\documentclass[10pt, conference]{IEEEtran}

\usepackage[margin = 0.7in]{geometry}
\usepackage{amsmath}
\usepackage{algpseudocode}
\usepackage{amssymb}
\usepackage{amsfonts}
\usepackage{amscd}
\usepackage{mathrsfs}
\usepackage{graphicx}
\usepackage[usenames,dvipsnames]{color}
\usepackage{bm}
\usepackage{url}
\usepackage{verbatim}
\usepackage[mathcal]{euscript}

\newtheorem{theorem}{Theorem}
\newtheorem{lemma}{Lemma}
\newtheorem{definition}{Definition}
\newtheorem{proposition}{Proposition}

\begin{document}

%\title{Storage Efficient Erasure Coded Archival of Versioned Data}
\title{Achievable Rates of Attack Detection Strategies in Echo-Assisted Communication}

\author{Mohit Goyal and J. Harshan\\
Department of Electrical Engineering,\\ 
Indian Institute of Technology Delhi, India.\\
%$^{\dagger}$Advanced Digital Sciences Center, Singapore, $^{\star}$University of Colorado, Colorado Springs, USA,\\ $^{*}$University of Illinois Urbana-Champaign, USA\\
Email: ee1150111@ee.iitd.ac.in and jharshan@ee.iitd.ac.in\\}

%\institute{$^{\dagger}$Advanced Digital Sciences Center, Singapore, $^{*}$University of Illinois Urbana-Champaign, USA\\
%Email:\{harshan.j, sychg\}@adsc.com.sg, yihchun@illinois.edu\\}

%

\maketitle

%********************************************************%
%
% Abstract
%
%**********************************************************%

\begin{abstract}
We consider an echo-assisted communication model wherein block-coded messages, when transmitted across several frames, reach the destination as multiple noisy copies. We address adversarial attacks on such models wherein a subset of the noisy copies are vulnerable to manipulation by an adversary. Particularly, we study a non-persistent attack model with the adversary attacking $50\%$ of the frames on the vulnerable copies in an i.i.d. fashion. We show that this adversarial model drives the destination to detect the attack locally within every frame, thereby resulting in degraded performance due to false-positives and miss-detection. Our main objective is to characterize the mutual information of this adversarial echo-assisted channel by incorporating the performance of attack-detection strategies. With the use of an imperfect detector, we show that the compound channel comprising the adversarial echo-assisted channel and the attack detector exhibits memory-property, and as a result, obtaining closed-form expressions on mutual information is intractable. To circumvent this problem, we present a new framework to approximate the mutual information by deriving sufficient conditions on the channel parameters and also the performance of the attack detectors. Finally, we propose two attack-detectors, which are inspired by traditional as well as neural-network ideas, and show that the mutual information offered by these detectors is close to that of the Genie detector for short frame-lengths. 
\end{abstract}

%\keywords{Adversarial communication, Achievable rates, Mutual information estimation}

%********************************************************%
%
% Introduction
%
%**********************************************************%
\section{Introduction} 
\label{sec:intro}

A number of wireless applications exists involving echo-assisted communication wherein messages transmitted by the source arrive at the destination as multiple noisy copies. Typical examples include communication over frequency-selective channels \cite{HSY}, relay networks \cite{NTW}, and multiple receive antennas \cite{mimo_ref}. In such scenarios, it is well known that suitably combining these copies can increase the effective signal-to-noise-ratio, thereby facilitating higher transmission rate. 

In this work, we consider attack models on echo-assisted communication wherein a subset of the copies collected at the destination might have been manipulated by an adversary. Attacks on only a subset of copies are attributed to practical limitations on the adversary to manipulate all the copies. For instance, in the case of frequency-selective channels with delay spreads, the adversary may have processing-delay constraints to manipulate the first copy, but not the subsequent ones \cite{HSY}. We study a specific adversarial attack referred to as the flipping attack \cite{DJLS} wherein the message bits of the attacked copy are flipped at 50\% rate independently. With such attacks, the dilemma at the destination is whether to use the vulnerable copies or discard them when recovering the messages. To gain insights on the attack model, we focus on the case of two received copies, out of which the second copy might have been manipulated by an adversary. Although adversarial models on binary channels have been studied by the information-theory community \cite{DJLS, BudJ}, flipping attacks on echo-assisted communication involving binary input and continuous output have not been studied hitherto. Henceforth, throughout the paper, we refer to the source and the destination as Alice and Bob, respectively.

\begin{center}
\begin{figure}[h]
\includegraphics[scale=0.44]{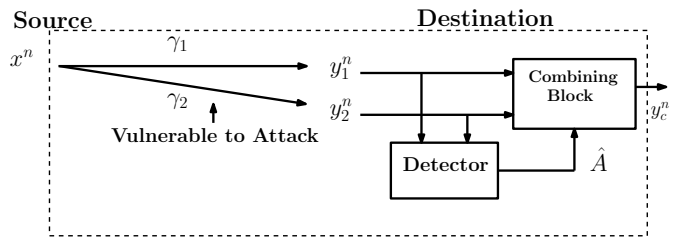}
\vspace{-0.5cm}
\caption{\label{fig:compound_channel} Compound channel comprising the source, adversarial echo-assisted channel, and the combining strategy, which is aided by the attack-detection block at the destination. In this work, we characterize the mutual information $I(x^{n};y^{n}_{c} ~|~ \hat{A})$ of the compound channel, where $x^{n} \in \{-1, +1\}^{n}$ is the input frame, $y^{n}_{1} \in \mathbb{R}^{n}$ and $y^{n}_{2} \in \mathbb{R}^{n}$ are the two received copies at the destination, $\hat{A}$ is the binary variable which represents the decision of the attack detector, and $y^{n}_{c} \in \mathbb{R}^{n}$ is the output of the combining block. } 
\end{figure}
\end{center}

%\begin{center}
%\begin{figure}[h]
%\includegraphics[scale=0.5]{ITW_frame_structure_new}
%\vspace{-0.2cm}
%\caption{\label{fig:frame_structure} Echo-assisted communication wherein a codeword, which is spread across multiple frames, arrives at the destination as two noisy copies. The adversarial model is such that the first copy is untouched by the attacker whereas the second copy is vulnerable to attack on $50\%$ of the frames chosen in an i.i.d. fashion. The destination must detect whether a frame is under attack before combining it with the first copy.} 
%\end{figure}
%\end{center}
\subsection{Motivation}
\label{subsec:motivation}

Consider an echo-assisted communication setting, as shown in Fig. \ref{fig:compound_channel}, wherein a binary codeword of large block-length is transmitted from Alice to Bob as a sequence of several frames, each of length $n$. Upon transmission of a frame, denoted by $x^{n}$, Bob receives two noisy copies of it, denoted by $y^{n}_{1} \in \mathbb{R}^{n}$ and $y^{n}_{2} \in \mathbb{R}^{n}$, in the presence of additive white Gaussian noise (AWGN). It is well known that appropriately combining these two copies can yield higher signal-to-noise-ratio at Bob, which in turn assists Alice to transmit at higher-rate than when only one of the copies is used to decode the codeword. The adversarial model in our setting is that the second copy is vulnerable to the flipping attack but not the first one. 
%If the attack is persistent, i.e., when executed across a large number of successive codewords, then Bob needs to detect the attack by observing $\{y^{n}_{1}, y^{n}_{2}\}$ over several codewords, and then feedback his decision on the combining strategy to Alice. Given that Bob offers two possible rates, one by combining the two copies, and the other by using only the first copy, Alice can design two codebooks and then use one of them based on the feedback from Bob. Intuitively, handling persistent attacks is straightforward for two reasons: (i) large number of samples are available at Bob to detect the attack, and (ii) Alice needs to use two traditional codebooks designed for Gaussian channels, however, with rates catering to the combining strategy at Bob.
Specifically, we consider a non-persistent attack model, wherein the second copy is vulnerable to the flipping attack on 50\% of the frames chosen at random in an i.i.d. fashion.\footnote{Persistent adversarial model, wherein all the frames of the second copy are under attack, is relatively straightforward to handle, as Bob may detect the attack accurately when the block-length of the code is large.} A conservative strategy to handle this adversarial setting is as follows: 
\begin{itemize}
\item Bob discards $y^{n}_{2}$ irrespective of the attack, and only uses $y^{n}_{1}$ to recover the message, i.e., $y^{n}_{c} = y^{n}_{1}$ as per Fig. \ref{fig:compound_channel}.
\item Alice uses a codebook designed for Gaussian channels to achieve the rate $I(x; y_{1})$, wherein $y_{1} = \gamma_{1}x + z_{1}$ such that $x \in \{-1, +1\}$, $z_{1} \sim \mathcal{N}(0, \sigma^2)$, and $\gamma_{1}$ is a constant known to both Alice and Bob.
\end{itemize}

\indent Keeping in view of the above conservative baseline, we are interested in designing a combining strategy at Bob which can assist Alice in transmitting at higher-rate than $I(x; y_{1})$. Towards achieving higher-rate, it is clear that Bob must first observe $y^{n}_{2}$, detect whether $y^{n}_{2}$ is attacked, and then decide to combine it with $y^{n}_{1}$ to recover the message. Since the frames are under attack in an i.i.d. fashion, Bob has to detect the attack locally by observing the $n$ samples of the frame, and this detection problem can be challenging especially when $n$ is small. Given that a practical detection strategy is typically imperfect, the combining strategy may lead to degraded performance either (i) when the flipping attack on $y^{n}_{2}$ is misdetected, or (ii) when a legitimate frame $y^{n}_{2}$ is categorized as under attack. While the performance of detectors can be evaluated by miss-detection and false-positive rates, these traditional metrics do not capture any rate-loss incurred by the source in aiding the detection strategy. As a result, there is a need to characterize the achievable rates of this adversarial echo-assisted channel in terms of the performance of the underlying attack-detectors.

\begin{table}
\begin{scriptsize}
\caption{Mutual Information Computation of Attack-Detectors in Echo-assisted Communication}
\vspace{-0.2cm}
\begin{center}
\begin{tabular}{|c|c|c|c|c|c|c|c|c|c|c|}
\hline \textbf{Operating Region of the Detector} & \textbf{Mutual Information}\\
& \textbf{Computation}\\
\hline $p_{md} = 0, p_{fa} = 0$ (Genie Detector) & Tractable\\
\hline $p_{md} = 0, p_{fa} = 1$ (Conservative Strategy) & Tractable\\
%\hline $p_{md} = 1, p_{fa} = 0$ & Tractable\\
%\hline $p_{md} = 1, p_{fa} = 1$ & Tractable\\
\hline $0 < p_{md} < 1, 0 < p_{fa} < 1$ & Intractable\\
\hline A special case of the regime & We propose \\
$0 < p_{md} < 1, 0 < p_{fa} < 1$ & an approximation in Theorem 1\\
\hline
\end{tabular}
\end{center}
\label{table:contribution_table}
\end{scriptsize}
\end{table}
\subsection{Contributions}

\begin{center}
\begin{figure}[h]
\includegraphics[scale=0.46]{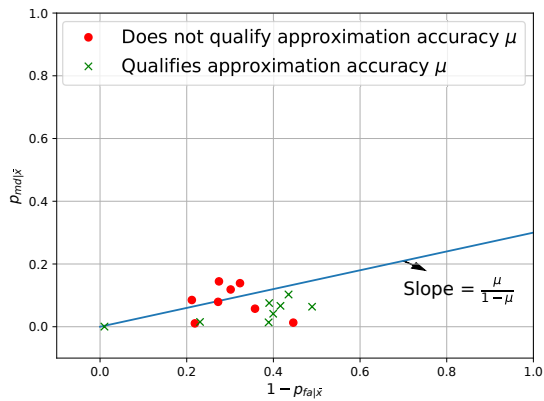}
\vspace{-0.3cm}
\caption{\label{fig:motivation}Plot of $\{(p_{md | \bar{x}}, 1- p_{fa | \bar{x}})\}$ of two detectors, where $p_{md | \bar{x}}$ and $p_{fa | \bar{x}}$ denote the miss-detection and false-positive rates conditioned on input codewords $x^{n} = \bar{x}$ for $n = 3$. We propose a framework to approximate the achievable rates of detectors which have $\{(p_{md | \bar{x}}, 1- p_{fa | \bar{x}}) ~|~ \bar{x} \in \{-1, 1\}^{n}\}$ below the line with slope $\frac{\mu}{1 - \mu}$ for some small $0 < \mu < 1$. To exemplify, given a small $\mu > 0$, our framework can approximate the rate of the detector marked with symbol $\times$ in green but not the one with $\circ$ in red.}
\end{figure}
\end{center}

The main contributions of this work are listed below:

\begin{itemize}
\item On the echo-assisted communication model discussed in Section \ref{subsec:motivation}, we quantify the performance of attack detectors by computing the mutual information of the compound channel, which comprises of the source, the adversarial echo-channel, and the combining strategy, which is guided by the attack detector at Bob, as shown in Fig. \ref{fig:compound_channel}. This way, we incorporate the traditional metrics of miss-detection rates, denoted by $p_{md}$, and false-positive rates, denoted by $p_{fa}$, indirectly into the mutual information of the compound channel. Henceforth, the mutual information of the compound channel, as shown in Fig. \ref{fig:compound_channel}, is referred to as the mutual information of the underlying attack-detector. 

\item Although the adversarial model has memory, we show that the compound channel involving a Genie detector (which corresponds to $p_{md} = 0$ and $p_{fa} = 0$) is memoryless by the virtue of perfect knowledge of the attack event at Bob. As a result, we show that computation of mutual information of Genie detectors is tractable (as listed in Table \ref{table:contribution_table}). However, it is well known that Genie detectors cannot be realized in practice especially when the frame-length is not sufficiently large. We show that the compound channel comprising a practical (imperfect) attack detector, such that $0 < p_{md} < 1, 0 < p_{fa} < 1$, continues to have memory, and this in addition to finitary constraint on the input alphabet \cite{HS}, renders mutual information computation intractable. To circumvent this issue, we provide a new framework to approximate the mutual information of imperfect detectors. Specifically, we provide sufficient conditions on (i) the miss-detection and false-positive rates of detectors (as shown in Fig. \ref{fig:motivation}), and (ii) on the channel parameters such that the proposed approximation holds (see Theorem \ref{thm1}).

\item In the last part of this work (see Section \ref{sec:det_techniques}), we propose two attack detectors, namely: (i) k Nearest Neighbor (KNN) estimator, which measures the mutual information between $y^{n}_{1}$ and $y^{n}_{2}$ to detect the flipping attack, and (ii) a Neural Network (NN) classifier, which uses two hidden layers to solve the detection problem as a supervised classification problem. We present our approximations on the mutual information of these detectors and show that the NN classifier is capable of accurately detecting the attacks on frames-lengths as short as $100$ and $40$ symbols at low signal-to-noise-ratio of $0$ dB and $5$ dB, respectively.

\end{itemize}

\emph{Notations}: For an $n$-dimensional random vector $y^{n} \in \mathbb{R}^{n}$ with joint probability distribution function $P(y^{n})$, its differential entropy, denoted by $h(y^{n})$, is represented as $-\mathbb{E}[\mbox{log}_{2}(P(y^{n}))]$, where the expectation is over $P(y^{n})$. A Gaussian random variable with zero mean and variance $\sigma^{2}$ is denoted by $\mathcal{N}(0, \sigma^{2})$. An $n \times n$ identity matrix, an $n$-length vector of zeros,  and an $n$-length vector of ones are denoted by $\mathbf{I}_{n}$, $\mathbf{0}_{n}$, and $\mathbf{1}_{n}$, respectively. For a given $n$-length vector, denoted by $y^{n}$, the notation $y^{n'}$ for $n' \leq n$, denotes the $n'$-length vector containing the first $n'$ components of $y^{n}$. The notation $\mbox{prob}(\cdot)$ denotes the usual probability operator.

\section{System Model}
\label{sec:system_model}

Alice transmits an $n$-length frame $x^{n} \in \{-1, +1\}^{n}$ such that the components of $x^{n}$ are i.i.d. over the Probability Mass Function (PMF) $\{\alpha, 1 - \alpha\}$ for some $0 < \alpha < 1$. Meanwhile, Bob receives two copies of $x^{n}$ over the Additive White Gaussian Noise (AWGN) channels as
\begin{equation}
\label{eq:signal_model}
y^{n}_{1} = \gamma_{1}x^{n} + z^{n}_{1} \in \mathbb{R}^{n} \mbox{ and } y^{n}_{2} = \gamma_{2}(b^{n} \circ x^{n}) + z^{n}_{2} \in \mathbb{R}^{n},
\end{equation}
where $\gamma_{1} \in \mathbb{R}$ and $\gamma_{2} \in \mathbb{R}$ are non-zero constants known to both Alice and Bob, $z^{n}_{1}$ and $z^{n}_{2}$ represent the additive white Gaussian noise vectors distributed as $\mathcal{N}(\mathbf{0}_{n}, \sigma^{2}\mathbf{I}_{n})$. We assume that $z^{n}_{1}$ and $z^{n}_{2}$ are statistically independent. Between the two copies, we assume that $y^{n}_{2}$ is vulnerable to the flipping attack, whereas $y^{n}_{1}$ is not. To model the flipping attack on $y^{n}_{2}$, we introduce Hadamard product, denoted by $\circ$, between $b^{n} \in \{-1, +1\}^{n}$ and $x^{n}$. When the frame is under attack, the components of $b^{n}$ are i.i.d. over the PMF $\{0.5, 0.5\}$, and are unknown to both Alice and Bob. However, without attack, $b^{n} = \mathbf{1}_{n}$. In this adversarial setting, the attacker executes the flipping attack on a frame chosen randomly in an i.i.d. fashion with probability $0.5$. By using $A = 0$ and $A = 1$ to denote the events of attack and no-attack, respectively, we have $\mbox{prob}(A = 0) = \mbox{prob}(A = 1) = 0.5$.

With no knowledge of $A$ at Bob, characterizing the mutual information (MI) of the adversarial channel is intractable due to the memory-property introduced by the attacker. However, when $A$ is perfectly known at Bob, we can compute the MI of the compound channel shown in Fig. \ref{fig:compound_channel}, wherein the underlying detector is the Genie detector, which assigns $\hat{A} = A$ for each frame.

\begin{proposition}
\label{prop_genie}
The MI of the compound channel involving the Genie detector is 
\begin{eqnarray}
\label{eq_mi_m_genie_prior}
\mathcal{M}^{Genie} & = & I(x; y_{c, na})\frac{n}{2} + I(x; y_{1})\frac{n}{2},
\end{eqnarray}
where $y_{c, na} = (|\gamma_{1}|^2 + |\gamma_{2}|^2)x + z_{c} \mbox{ and } y_{1} = \gamma_{1}x + z_{1}$ are the scalar channels such that $x \in \{-1, +1\}$ with PMF $\{\alpha, 1 - \alpha\}$, and the additive noise $z_{c}  = \gamma_{1}z_{1} + \gamma_{2}z_{2}$ is distributed as $\mathcal{N}(0, \sigma^{2}_{eq})$ with $\sigma^{2}_{eq} = (|\gamma_{1}|^{2} + |\gamma_{2}|^{2})\sigma^{2}$.
\end{proposition}
\begin{IEEEproof}
The average MI offered by the compound channel comprising the Genie detector is 
\begin{eqnarray}
\label{eq:Genie_ergodic}
\mathcal{M}^{Genie} & = & I(x^{n}; y^{n}_{1}, y^{n}_{2} ~|~ A = 0)\mbox{prob}(A = 0)\nonumber \\
& & + ~I(x^{n}; y^{n}_{1}, y^{n}_{2} ~|~ A = 1)\mbox{prob}(A = 1),\nonumber \\
& = & \frac{1}{2} I(x^{n}; y^{n}_{1}, y^{n}_{2} ~|~ A = 0)\nonumber \\
& & + ~ \frac{1}{2}I(x^{n}; y^{n}_{1}, y^{n}_{2} ~|~ A = 1).
\end{eqnarray}
When $A = 1$, each bit of $x^{n}$ on the second copy is flipped by the attacker with probability $0.5$ in an i.i.d. fashion, and as a result, it is straightforward to prove that $I(x^{n}; y^{n}_{2} ~|~ A = 1) = 0$. As a consequence, we have 
\begin{eqnarray}
\label{eq:Genie_ergodic_1}
I(x^{n}; y^{n}_{1}, y^{n}_{2} ~|~ A = 1) = I(x^{n}; y^{n}_{1}) = nI(x; y_{1}),
\end{eqnarray}
where the last equality is applicable due to the memoryless property of the channel on the first copy. This implies that discarding $y^{n}_{2}$ is the optimal strategy at Bob when $A = 1$. On the other hand, when $A = 0$, the mutual information of the compound channel is given by
\begin{eqnarray}
\label{eq:Genie_ergodic_2}
I(x^{n}; y^{n}_{1}, y^{n}_{2} ~|~ A = 0) & = & I(x^{n}; y^{n}_{1}, y^{n}_{2} ~|~ b^{n} = \mathbf{1}_{n}),\nonumber\\
& = & I(x^{n}; y^{n}_{c, na}),\nonumber\\
& = & nI(x; y_{c, na}),
\end{eqnarray}
where $y^{n}_{c, na}$ in the second equality is obtained by combining $y^{n}_{1}$ and $y^{n}_{2}$ as $y^{n}_{c, na} = \gamma_{1}y^{n}_{1} + \gamma_{2}y^{n}_{2} = (|\gamma_{1}|^2 + |\gamma_{2}|^2)x^{n} + z^{n}_{c},$ such that the additive noise vector $z^{n}_{c}  = \gamma_{1}z^{n}_{1} + \gamma_{2}z^{n}_{2}$ is distributed as $\mathcal{N}(\mathbf{0}_{n}, \sigma^{2}_{eq}\mathbf{I}_{n})$, where $\sigma^{2}_{eq} = (|\gamma_{1}|^{2} + |\gamma_{2}|^{2})\sigma^{2}$. It is straightforward to verify that $I(x^{n}, y^{n}_{c, na}) = I(x^{n}; y^{n}_{1}, y^{n}_{2} ~|~ b^{n} = \mathbf{1}_{n})$, which implies that the combining strategy is optimal without the attack. Note that the last equality is applicable by using the memoryless nature of the channel, attributed to the perfect knowledge of $A$ at Bob. Finally, by using \eqref{eq:Genie_ergodic_1} and \eqref{eq:Genie_ergodic_2} in \eqref{eq:Genie_ergodic}, we get the expression for mutual information in \eqref{eq_mi_m_genie_prior}. This completes the proof. 
\end{IEEEproof}

\begin{figure*}
\begin{equation}
\label{eq:pdf_y_no_attack}
P(y_{c, na}) = \frac{1}{\sqrt{2\pi \sigma^{2}_{eq}}}\left(\alpha e^{-\frac{(y_{c, na} - (|\gamma_{1}|^2 + |\gamma_{2}|^2))^{2}}{2\sigma^{2}_{eq}}} + (1-\alpha) e^{-\frac{(y_{c, na} + (|\gamma_{1}|^2 + |\gamma_{2}|^2))^{2}}{2\sigma^{2}_{eq}}}\right)\\ 
\end{equation}
\hrule
\end{figure*}

Since $x$ takes values from a finite input alphabet, $\mathcal{M}^{Genie}$ in \eqref{eq_mi_m_genie_prior} can be numerically computed as a function of the input PMF $\{\alpha, 1 - \alpha\}$, constants $\gamma_{1}$ and $\gamma_{2}$, and $\sigma^{2}$ \cite{HS}. Specifically, $I(x; y_{c, na})$ is given by
\begin{equation}
\label{eq:mi_no_attack}
I(x; y_{c, na}) = h(y_{c, na}) - h(y_{c, na} |x),
\end{equation}
where $h(y_{c, na}) = -\mathbb{E}[\mbox{log}_{2}(P(y_{c, na}))]$ such that $P(y_{c, na})$ is as given in \eqref{eq:pdf_y_no_attack}. The conditional entropy $h(y_{c, na}|x)$ can be computed using the distribution $P(y_{c, na} | x = \beta)$ given by 
%\begin{equation*}
%h(y_{c, na}|x) = \alpha h(y_{c, na}|x = 1) + (1-\alpha) h(y_{c, na}|x = -1)
%\end{equation*}
%where $h(y_{c, na}|x = \beta) = -\mathbb{E}[\mbox{log}_{2}(P(y_{c, na}|x = \beta))]$ and 
\begin{eqnarray*}
P(y_{c, na} | x = \beta) = \frac{1}{\sqrt{2\pi \sigma^{2}_{eq}}} e^{-\frac{(y_{c, na} - \beta(|\gamma_{1}|^2 + |\gamma_{2}|^2))^{2}}{2\sigma^{2}_{eq}}},
\end{eqnarray*}
for $\beta \in \{-1, +1\}$.
%Using the above distributions, we can numerically compute $I(x; y_{c, na})$ in \eqref{eq:mi_no_attack}. 
Similarly, we can also compute $I(x; y_{1})$. 

Although the mutual information of Genie detectors can be computed based on Proposition \ref{prop_genie}, it is well known that practical detectors not perfect. Therefore, in the next section, we address the challenges involved in computing the MI of (imperfect) practical attack-detectors. 
%\subsection{Achievable rate of attack-ignorant receiver}
%
%When the receiver, referred to as Bob, is not aware of the flipping attack, he will combine both the copies every frame, thereby obtaining 
%\begin{equation*}
%y^{n}_{1, 2} = h^{*}_{1}y^{n}_{1} + h^{*}_{2}y^{n}_{2}
%\end{equation*}
%When there is an attack, $y^{n}_{1, 2, a}$ is of the form
%\begin{equation*}
%y^{n}_{1, 2, a} = (|h_{1}|^{2} + b^{n}|h_{1}|^{2}) \odot x^{n} + z^{n}_{1,2}
%\end{equation*}
%where the operator $\odot$ represents component-wise multiplication. However, without attack, $y^{n}_{1, 2}$ is of the form
%\begin{equation*}
%y^{n}_{1, 2} = (|h_{1}|^{2} + |h_{1}|^{2}) x^{n} + z^{n}_{1,2}
%\end{equation*}
%Since the receiver does not know about the attack, the probability distribution on the combined samples is given by
%\begin{equation*}
%P(y^{n}_{ignorant}) = \frac{1}{2}(P(y^{n}_{1, 2}) + P(y^{n}_{1, 2, a}))
%\end{equation*}
%Let us first study the achievable rate $I(y^{n}_{ignorant}; x^{n})$.

\section{Mutual Information with Practical Detection Strategy}

\begin{figure}
\includegraphics[scale=0.45]{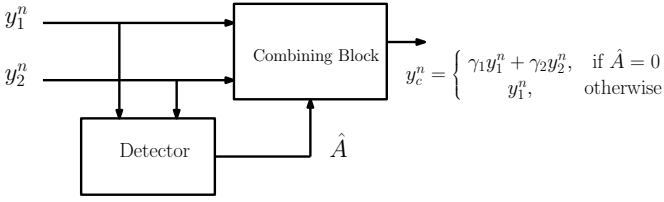}
\caption{\label{fig:pract_detection_strategy}Depiction of the combining strategy with a practical detection algorithm.}
\end{figure}

We consider a practical attack-detection strategy, as shown in Fig. \ref{fig:pract_detection_strategy}, which uses the received samples $\{y^{n}_{1}, y^{n}_{2}\}$ to detect the flipping attack on every frame. Based on the detector's output, represented by the variable $\hat{A} \in \{0, 1\}$, Bob decides either to combine $y^{n}_{1}$ and $y^{n}_{2}$, or discard $y^{n}_{2}$. Note that this detector is typically imperfect, and as a result, it has its associated miss-detection and false-positive rates, defined as $p_{md} \triangleq \mbox{prob}(\hat{A} = 0 ~|~ A = 1)$ and $p_{fa} \triangleq \mbox{prob}(\hat{A} = 1 ~|~ A = 0)$, respectively. When the detector outputs $\hat{A} = 1$, Bob drops the samples $y^{n}_{2}$, and only uses the samples $y^{n}_{1}$ to recover the message. On the other hand, when the detector outputs $\hat{A} = 0$, Bob combines $y^{n}_{1}$ and $y^{n}_{2}$ to obtain $y^{n}_{c} = \gamma_{1}y^{n}_{1} + \gamma_{2}y^{n}_{2}$ and then uses it to recover the message. 

In the event of miss-detection, i.e., when $A = 1$ and $\hat{A} = 0$, we know that $b^{n} \in \{-1, +1\}^{n}$ is random and unknown to Bob. Therefore, $y^{n}_{c}$ is denoted as $y^{n}_{c, a}$, and is given by
\begin{eqnarray}
\label{eq:rx_combine_when_attack}
y^{n}_{c, a} = (|\gamma_{1}|^2 + b^{n}|\gamma_{2}|^2) \circ x^{n} + z^{n}_{c}.
\end{eqnarray}
However, when $A = 0$ and $\hat{A} = 0$, we have $b^{n} = \textbf{1}_{n}$, and therefore, $y^{n}_{c}$ is denoted as $y^{n}_{c, na}$, and is given by
\begin{eqnarray}
\label{eq:rx_combine_when_no_attack}
y^{n}_{c, na} = (|\gamma_{1}|^2 + |\gamma_{2}|^2)x^{n} + z^{n}_{c}.
\end{eqnarray}
The MI of this detection strategy, denoted by $\mathcal{M}^{non-Genie}_{p_{md}, p_{fa}}$, is 
\begin{eqnarray}
\label{eq:rate_non_genie}
\mathcal{M}^{non-Genie}_{p_{md}, p_{fa}} & = & I(x^{n}; y^{n}_{c} ~|~ \hat{A} = 0)\mbox{prob}(\hat{A} = 0) \nonumber\\
& & + ~I(x^{n}; y^{n}_{1})\mbox{prob}(\hat{A} = 1),
%& = & I(x^{n}; y^{n}_{1, 2} ~|~ \hat{A} = 0)\mbox{prob}(\hat{A} = 0) + I(x^{n}; y^{n}_{1})\mbox{prob}(\hat{\mathcal{A}} = 1)\nonumber\\
\end{eqnarray}
where $\mbox{prob}(\hat{A} = 0) = \frac{1}{2}(1 + p_{md} - p_{fa})$ and $\mbox{prob}(\hat{A} = 1) = \frac{1}{2}(1 - p_{md} + p_{fa})$.

To compute $\mathcal{M}^{non-Genie}_{p_{md}, p_{fa}}$, we have to compute $I(x^{n}; y^{n}_{c} ~|~ \hat{A} = 0)$ for a given frame-length $n$. However, this needs us to evaluate the differential entropy of the probability distribution function $P(y^{n}_{c} ~|~ \hat{A} = 0)$ given in \eqref{pdf_y_combine_delta_A_0_expanded}. Since the input alphabet is finite in size, the corresponding differential entropy can only be computed using numerical methods, and as a result, computing $\mathcal{M}^{non-Genie}_{p_{md}, p_{fa}}$ is intractable for sufficiently large $n$ (of the order of hundreds). In a nutshell, the above computational issue is because the equivalent channel when $\hat{A} = 0$ is not memoryless. To circumvent this problem, we show that the MI value $\mathcal{M}^{non-Genie}_{p_{md}, p_{fa}}$ of some detectors can be computed using an approximation under special conditions on $p_{md}$ and $p_{fa}$.
%obtaining $I(x^{n}; y^{n}_{1})$ is straightforward (as discussed when computing $\mathcal{M}^{Genie}$ in Section \ref{sec:system_model}), whereas obtaining $I(x^{n}; y^{n}_{c} ~|~ \hat{A} = 0)$ is challenging for large $n$; the probability distribution function $P(y^{n}_{c} ~|~ \hat{A} = 0)$ needed to compute $I(x^{n}; y^{n}_{c} ~|~ \hat{A} = 0)$, is of the form \eqref{pdf_y_combine_delta_A_0} this is because the event $\hat{A} = 0$ involves miss-detection, and as a result, the equivalent channel is not memoryless. 

\begin{figure*}
\begin{eqnarray}
P(y^{n}_{c} ~|~ \hat{A} = 0) & = & \frac{P(y^{n}_{c} | A = 1, \hat{A} = 0)\mbox{prob}(A = 1, \hat{A} = 0) + P(y^{n}_{c} | A = 0, \hat{A} = 0)\mbox{prob}(A = 0, \hat{A} = 0)}{\mbox{prob}(\hat{A} = 0)}\nonumber \\
\label{pdf_y_combine_delta_A_0}
& = & \frac{P(y^{n}_{c, a})p_{md} + P(y^{n}_{c, na})(1-p_{fa})}{p_{md} + 1 - p_{fa}} \\
\label{pdf_y_combine_delta_A_0_expanded}
P(y^{n}_{c} ~|~ \hat{A} = 0) & = & \frac{p_{md}}{(2 \pi \sigma^{2}_{eq})^{\frac{n}{2}} (p_{md} + 1 - p_{fa})} \frac{1}{2^{n}} \sum_{x^{n} = \bar{x}} \mbox{prob}(x^{n} = \bar{x}) \left(\sum_{b^{n} = \bar{b}} e^{- \frac{||y^{n}_{c} - (|\gamma_{1}|^{2}  + \bar{b}|\gamma_{2}|^2)\circ \bar{x}||^{2}_{F}}{2\sigma^{2}_{eq}}}\right) \\
& & + ~ \frac{1 - p_{fa}}{(2 \pi \sigma^{2}_{eq})^{\frac{n}{2}}(p_{md} + 1 - p_{pfa})} \sum_{x^{n} = \bar{x}} \mbox{prob}(x^{n} = \bar{x}) e^{- \frac{||y^{n}_{c} - (|\gamma_{1}|^{2}  + |\gamma_{2}|^2) \bar{x}||^{2}_{F}}{2\sigma^{2}_{eq}}} \nonumber
\end{eqnarray}
\hrule
\end{figure*}

%Towards computing $\mathcal{M}^{non-genie}$, obtaining $I(x^{n}; y^{n}_{1})$ is straightforward, whereas obtaining $I(x^{n}; y^{n}_{c} ~|~ \hat{\mathcal{A}} = 0)$ is challenging; this is because the event $\hat{\mathcal{A}} = 0$ also involves misdetections. Furthermore, in order to achieve \eqref{eq:rate_non_genie}, Alice would need to design two codebooks: (1) one aligned to $I(x^{n}; y^{n}_{c} ~|~ \hat{\mathcal{A}} = 0)$, and (ii) the other aligned to $I(x^{n}; y^{n}_{1})$. Since the distribution of $y^{n}_{c}$ is multi-modal, designing a codebook that achieves this rate is challenging. The following theorem shows that when the threshold value of the detection strategy is varied to have $p_{md} = 0$, then the corresponding mutual-information value $\mathcal{M}^{non-genie}$ can be computed and also be achieved using a codebook designed for Gaussian channels with binary input alphabet. 
The following sequence of definitions and lemmas are useful to present our results on approximation in Theorem \ref{thm1}. 

\begin{definition}
\label{def2}
For $0 \leq x, y \leq 1$, let a set $\mathcal{R}_{\mu}$, for some negligible $\mu > 0$, be defined as
\begin{equation*}
\mathcal{R}_{\mu} \triangleq \left\lbrace(x, y) ~|~ y \leq \frac{\mu}{1 - \mu}x\right\rbrace.
\end{equation*}
\end{definition}

\begin{definition}
\label{def3}
For a given attack-detector, we define its performance profile as
\begin{equation*}
\mathcal{P} \triangleq \left\lbrace (p_{md | \bar{x}}, 1 - p_{fa | \bar{x}}) ~|~ \forall ~\bar{x} \in \{-1, 1\}^{n} \right\rbrace,
\end{equation*}
where $p_{md | \bar{x}} = \mbox{prob}(\hat{A} = 0 | A = 1, x^{n} = \bar{x})$ and $p_{fa | \bar{x}} = \mbox{prob}(\hat{A} = 1 | A = 0, x^{n} = \bar{x})$. 
\end{definition}

\begin{definition}
\label{def1}
For a given $\bar{x} \in \{-1, 1\}^{n}$, let $\mathcal{S}_{\bar{x}} = \{(|\gamma_{1}|^{2}  + \bar{b} |\gamma_{2}|^2) \circ \bar{x} ~|~ \forall ~\bar{b} = \{-1, 1\}^{n}\}$ denote an $n$-dimensional discrete constellation in $\mathbb{R}^{n}$ obtained by using $\bar{b}$ over $\{-1, +1\}^{n}$. On $\mathcal{S}_{\bar{x}}$, we define,
\begin{itemize}
\item $d^{2}_{min}(y^{n}, \mathcal{S}_{\bar{x}}) = \displaystyle \min_{s^{n} \in \mathcal{S}_{\bar{x}}}   ||y^{n} - s^{n}||^{2}_{F}$ 
\item $d^{2}_{max}(y^{n}, \mathcal{S}_{\bar{x}}) = \displaystyle \max_{s^{n} \in \mathcal{S}_{\bar{x}}}  ||y^{n} - s^{n}||^{2}_{F}$
\item $d^{2}_{max}(\mathcal{S}_{\bar{x}}) = \displaystyle \max_{s^{n}_{1}, s^{n}_{2} \in \mathcal{S}_{\bar{x}}}  ||s^{n}_{1} - s^{n}_{2}||^{2}_{F},$
\end{itemize}
where $y^{n} \in \mathbb{R}^{n}$ and $||\cdot||^{2}_{F}$ denotes the squared Euclidean distance. 
\end{definition}

\begin{lemma}
\label{lemma1}
If $a, b, \mu$ are such that $0 \leq a \leq 2b$ and $\mu > 0$ is a negligible number, then we have $\mu a + (1 - \mu) b \approx b.$
\end{lemma}
\begin{IEEEproof}
The convex combination $\mu a + (1 - \mu) b$ can be written as $b - \mu(b-a)$. This implies that $\mu a + (1 - \mu) b = b - \lambda$, where $0 \leq \lambda \leq \mu b$ when $a \leq b$, and $-\mu b \leq \lambda < 0$ when $b < a \leq 2b$. Since $\mu$ is negligible, $b - \lambda \approx b$ for every $b \geq 0$.
\end{IEEEproof}

Since the accuracy of the approximation depends on $\mu$, we henceforth denote $\approx$ by $\approx_{\mu}$.

\begin{lemma}
\label{lemma2}
If $\gamma_{1}$, $\gamma_{2}$ and $\sigma^{2}_{eq}$ are such that $d^{2}_{max}(\mathcal{S}_{\bar{x}}) \leq 2\mbox{log}_{e}(2)\sigma^{2}_{eq}$ for each $\bar{x} \in \{-1, +1\}^{n},$ then we have
\begin{eqnarray}
\label{pdf_bound1}
P(y^{n}_{c, a} = y^{n} | \bar{x}) & \leq & 2P(y^{n}_{c, na} = y^{n} | \bar{x}),\\
\label{pdf_bound2}
P(y^{n}_{c, a} = y^{n}) & \leq & 2P(y^{n}_{c, na} = y^{n}),
\end{eqnarray}
for every $y^{n} \in \mathbb{R}^{n}$.
\end{lemma}
\begin{IEEEproof}
%We have omitted the proof due to lack of space. We refer the reader to the proof of Lemma 2 in \cite{harshan_mohit}.
We only show the applicability of \eqref{pdf_bound1}. Since $P(y^{n}_{c, a} = y^{n})$ can be written as a weighted sum of $P(y^{n}_{c, a} = y^{n} | \bar{x})$ over all $\bar{x}$, \eqref{pdf_bound1} can be used to show the applicability of \eqref{pdf_bound2}. Given $x^{n} = \bar{x}$, the $n$-dimensional distribution of $y^{n}_{c, a}$ is given by $P(y^{n}_{c, a} | \bar{x}) = \frac{1}{(2 \pi \sigma^{2}_{eq})^{\frac{n}{2}}} \frac{1}{2^{n}} \sum_{b^{n} = \bar{b}} e^{- \frac{||y^{n}_{c, a} - (|\gamma_{1}|^{2}  + \bar{b}|\gamma_{2}|^2)\circ \bar{x}||^{2}_{F}}{2\sigma^{2}_{eq}}}.$ When evaluated at $y^{n} \in \mathbb{R}^{n}$, we can upper bound the above term as 
\begin{equation}
\label{eq:term1}
P(y^{n}_{c, a}  = y^{n}| \bar{x}) \leq \frac{1}{(2 \pi \sigma^{2}_{eq})^{\frac{n}{2}}} e^{- \frac{d^{2}_{min}(y^{n}, \mathcal{S}_{\bar{x}})}{2\sigma^{2}_{eq}}},
\end{equation}
where $d^{2}_{min}(y^{n}, \mathcal{S}_{\bar{x}})$ is as given in Definition \ref{def1}.
Meanwhile, the $n$-dimensional distribution of $y^{n}_{c, na}$ is given by
\begin{eqnarray}
\label{eq:term2}
P(y^{n}_{c, na} = y^{n}| \bar{x}) & = & \frac{1}{(2 \pi \sigma^{2}_{eq})^{\frac{n}{2}}} e^{- \frac{||y^{n} - (|\gamma_{1}|^{2}  + |\gamma_{2}|^2)\bar{x}||^{2}_{F}}{2\sigma^{2}_{eq}}}, \nonumber \\
& \geq & \frac{1}{(2 \pi \sigma^{2}_{eq})^{\frac{n}{2}}} e^{- \frac{d^{2}_{max}(y^{n}, \mathcal{S}_{\bar{x}})}{2\sigma^{2}_{eq}}}, \nonumber \\
& \geq & \frac{1}{(2 \pi \sigma^{2}_{eq})^{\frac{n}{2}}} e^{- \frac{d^{2}_{min}(y^{n}, \mathcal{S}_{\bar{x}}) + d^{2}_{max}(\mathcal{S}_{\bar{x}})}{2\sigma^{2}_{eq}}}
\end{eqnarray}
where the first inequality holds since $(|\gamma_{1}|^{2}  + |\gamma_{2}|^2)\bar{x} \in \mathcal{S}_{\bar{x}}$. The second inequality holds because of triangle inequality. Finally, if $d^{2}_{max}(\mathcal{S}_{\bar{x}}) \leq 2\mbox{log}_{e}(2)\sigma^{2}_{eq}$ for each $\bar{x} \in \{-1, +1\}^{n},$ then \eqref{eq:term2} can be further lower bounded as 
\begin{eqnarray}
\label{eq:term3}
P(y^{n}_{c, na} = y^{n}| \bar{x}) & \geq & \frac{1}{(2 \pi \sigma^{2}_{eq})^{\frac{n}{2}}} e^{- \frac{d^{2}_{min}(y^{n}, \mathcal{S}_{\bar{x}}) + 2 \small{\mbox{log}}_{e}(2)\sigma^{2}_{eq}}{2\sigma^{2}_{eq}}},\nonumber \\
& = & \frac{1}{2(2 \pi \sigma^{2}_{eq})^{\frac{n}{2}}} e^{- \frac{d^{2}_{min}(y^{n}, \mathcal{S}_{\bar{x}})}{2\sigma^{2}_{eq}}},\\
& \geq & \frac{1}{2} P(y^{n}_{c, a} = y^{n}| \bar{x}),
\end{eqnarray}
where the last inequality is due to the bound in \eqref{eq:term1}. This implies that $P(y^{n}_{c, a} = y^{n} | \bar{x}) \leq 2P(y^{n}_{c, na} = y^{n} | \bar{x})$ for each $y^{n}$. This completes the proof.
\end{IEEEproof}

Using the results of Lemma $\ref{lemma1}$ and Lemma $\ref{lemma2}$, we are now ready to present our result on approximation.

\begin{theorem}
\label{thm1}
If $\gamma_{1}$, $\gamma_{2}$ and $\sigma^{2}_{eq}$ are such that $d^{2}_{max}(\mathcal{S}_{\bar{x}}) \leq 2\mbox{log}_{e}(2)\sigma^{2}_{eq}$ for each $\bar{x} \in \{-1, +1\}^{n},$ and if the detection strategy is such that $\mathcal{P} \subseteq \mathcal{R}_{\mu}$, for a fixed small $\mu > 0$, then we have $\mathcal{M}^{non-Genie}_{p_{md}, p_{fa}} \approx_{\mu, pdf} \mathcal{M}^{approx}_{p_{fa}}$, where 
\begin{equation}
\label{eq:rate_non_genie_pmd_0}
\mathcal{M}^{approx}_{p_{fa}} = \frac{n}{2}I(x; y_{c, na})(1-p_{fa}) + \frac{n}{2}I(x; y_{1})(1+ p_{fa}),
\end{equation}
and the notation $\approx_{\mu, pdf}$ captures the notion that the approximation on MI is a result of approximating the underlying distributions using $\approx_{\mu}$.
\end{theorem}
\begin{IEEEproof}
Based on the expression of $\mathcal{M}^{non-Genie}_{p_{md}, p_{fa}}$ in \eqref{eq:rate_non_genie}, it is straightforward to show that $I(x^{n}; y^{n}_{1}) = nI(x; y_{1})$. In this proof, we only address the computation of $I(x^{n}; y^{n}_{c} ~|~ \hat{A} = 0)$. From first principles, we have
\begin{equation*}
I(x^{n}; y^{n}_{c} ~|~ \hat{A} = 0) = h(y^{n}_{c} ~|~ \hat{A} = 0) - h(y^{n}_{c} ~|~ x^{n}, \hat{A} = 0), 
\end{equation*}
where $h(y^{n}_{c} ~|~ \hat{A} = 0)$ can be obtained using $P(y^{n}_{c} ~|~ \hat{A} = 0)$ as 
\begin{equation*}
h(y^{n}_{c} ~|~ \hat{A} = 0) = -\mathbb{E}\left[\mbox{log}_{2}\left(P(y^{n}_{c} ~|~ \hat{A} = 0)\right)\right],
\end{equation*}
where $P(y^{n}_{c} ~|~ \hat{A} = 0)$ is as given in \eqref{pdf_y_combine_delta_A_0}.
When the attack-detection technique operates at $\mathcal{P} \subseteq \mathcal{R}_{\mu}$, then we can show that $(p_{md}, p_{fa}) \in \mathcal{R}_{\mu}$, where $p_{fa} = \mathbb{E}[p_{fa|\bar{x}}]$ and $p_{md} = \mathbb{E}[p_{md|\bar{x}}]$ such that the expectation is over $x^{n}$. By applying the results of Lemma \ref{lemma1} and Lemma \ref{lemma2} on \eqref{pdf_y_combine_delta_A_0}, we get 
\begin{equation*}
P(y^{n}_{c} ~|~ \hat{A} = 0) \approx_{\mu} P(y^{n}_{c, na}).
\end{equation*}
The above approximation holds because $\frac{p_{md}}{p_{md} + 1 - p_{fa}}$ plays the role of $\mu$ in Lemma \ref{lemma1}, and the condition $a \leq 2b$ of Lemma \ref{lemma1} is satisfied because of \eqref{pdf_bound2} in Lemma \ref{lemma2}. As a result $h(y^{n}_{c} ~|~ \hat{A} = 0) \approx_{\mu, pdf} -\mathbb{E}[\mbox{log}_{2}(P(y^{n}_{c, na}))]$. Furthermore, since each component of $y^{n}_{c, na}$ is independent across $n$, we have
\begin{equation}
\label{eq:appro_diff_entropy}
h(y^{n}_{c} ~|~ \hat{A} = 0) \approx_{\mu, pdf} h(y^{n}_{c, na}) = nh(y_{c, na}),
\end{equation}
where $h(y_{c, na}) = -\mathbb{E}[\mbox{log}_{2}(P(y_{c, na}))]$ such that $P(y_{c, na})$ is given by \eqref{eq:pdf_y_no_attack}. Similarly, the conditional differential entropy $h(y^{n}_{c} ~|~ \hat{A} = 0, x^{n})$ is given by 
\begin{equation}
\label{eq:diff_cond_entropy}
h(y^{n}_{c} ~|~ \hat{A} = 0, x^{n}) = \sum_{x^{n} = \bar{x}} p(\bar{x} | \hat{A} = 0)h(y^{n}_{c} ~|~ \hat{A} = 0, x^{n} = \bar{x}),
\end{equation}
where $p(\bar{x} | \hat{A} = 0) \triangleq \mbox{prob}(x^{n} = \bar{x}|\hat{A} = 0)$ and $h(y^{n}_{c} ~|~ \hat{A} = 0, x^{n} = \bar{x}) = -\mathbb{E}[\mbox{log}_{2}(P(y^{n}_{c} ~|~ \hat{A} = 0, x^{n} = \bar{x}))]$ such that $P(y^{n}_{c} ~|~ \hat{A} = 0, x^{n} = \bar{x})$ can be written as
\begin{eqnarray}
\label{eq:cond_pdf}
\frac{P(y^{n}_{c, a}~|~x^{n} = \bar{x})p_{md | \bar{x}} + P(y^{n}_{c, na}~|~x^{n} = \bar{x})(1-p_{fa | \bar{x}})}{p_{md | \bar{x}} + 1 - p_{fa | \bar{x}}}.
\end{eqnarray}
To arrive at \eqref{eq:cond_pdf}, we assume that $A$ and $x^{n}$ are statistically independent. Again, applying the results of Lemma \ref{lemma1} and Lemma \ref{lemma2} on \eqref{eq:cond_pdf}, we have the approximation
\begin{equation*}
P(y^{n}_{c} ~|~ \hat{A} = 0, x^{n} = \bar{x}) \approx_{\mu} P(y^{n}_{c, na}~|~x^{n} = \bar{x}),
\end{equation*}
for every $x^{n} = \bar{x}$. As a result, we have $h(y^{n}_{c} ~|~ \hat{A} = 0, x^{n} = \bar{x}) \approx_{\mu, pdf} h(y^{n}_{c, na} ~|~ x^{n} = \bar{x})$. Finally, using the above expression in \eqref{eq:diff_cond_entropy}, we get 
\begin{small}
\begin{eqnarray}
\label{eq:appro_diff_cond_entropy}
h(y^{n}_{c} ~|~ \hat{A} = 0, x^{n}) & \approx_{\mu, pdf} & \sum_{x^{n} = \bar{x}} p(\bar{x} | \hat{A} = 0)h(y^{n}_{c, na} ~|~ x^{n} = \bar{x})\nonumber \\
%& = & \sum_{x^{n} = \bar{x}} p(\bar{x} | \hat{A} = 0)h(z^{n}_{c})\nonumber \\
& = & h(z^{n}_{c}) = nh(y_{c, na} ~|~ x),
\end{eqnarray}
\end{small}
\noindent where the last equality is due to i.i.d. nature of $z_{c}^{n}$. Overall, using \eqref{eq:appro_diff_cond_entropy} and \eqref{eq:appro_diff_entropy} in \eqref{eq:rate_non_genie}, we get the expression in \eqref{eq:rate_non_genie_pmd_0}. 
\end{IEEEproof}
$~~$\\
%To study the achievable rate of a practical detection strategy, we need to compute $\mathcal{M}^{non-Genie}_{p_{md}, p_{fa}}$ in \eqref{eq:rate_non_genie}, which is a function of (i) $\{\alpha, 1-\alpha\}$, the PMF on $x$, (ii) the miss-detection rate $p_{md}$, (iii) the false-positive rate $p_{fa}$, and (iv) the channel parameters $\gamma_{1}, \gamma_{1}$ and $\sigma^{2}$. However, 
\indent The proposed sufficient condition on the performance profile of attack detectors is also depicted in Fig. \ref{fig:motivation}. Due to intractability in evaluating $\mathcal{M}^{non-Genie}_{p_{md}, p_{fa}}$, Theorem \ref{thm1} approximates the MI of a special class of detectors when (i) the detectors operate in the region $\mathcal{P} \subseteq \mathcal{R}_{\mu}$, and (ii) the channel parameters $\gamma_{1}, \gamma_{2}, \sigma^{2}$ satisfy Lemma \ref{lemma2}. For such a class of detectors, the MI $\mathcal{M}^{approx}_{p_{fa}}$, given in \eqref{eq:rate_non_genie_pmd_0} is now easy to evaluate since $I(x; y_{c, na})$ and $I(x; y_{1})$ can be computed using standard numerical methods \cite{HS}. Note that the Genie detector trivially belongs to this special class, and as a result, \eqref{eq:rate_non_genie_pmd_0} is upper bounded by $\mathcal{M}^{Genie}$ in \eqref{eq_mi_m_genie_prior}. Also note that \eqref{eq:rate_non_genie_pmd_0} is lower bounded by $nI(x; y_{1})$, which is the MI offered by the conservative strategy of unconditionally dropping $y^{n}_{2}$ when recovering the message.

\section{Experiment Results}
\label{sec:det_techniques} 

To conduct experiments on the performance of attack detection in echo-assisted communication, we use the system model in Section \ref{sec:system_model} with $\alpha = 0.5$, $\gamma_{1} = \gamma_{2} = 1$, and $\mbox{SNR} = 10\mbox{log}_{10}(\frac{1}{\sigma^{2}}) \in \{0, 5, 10, 15\}$ in dB. We propose the following two detectors which are designed to detect the flipping attack by using the first $n'$ samples of the received frames, namely $\{y^{n'}_{1}, y^{n'}_{2}\}$, for some $n' \leq n$. 

\textit{1) k Nearest-Neighbor (KNN) MI Estimation:} Based on the attack model in Section \ref{sec:system_model}, we observe that $I(y_{1}; y_{2} ~|~ A = 0) > I(y_{1}; y_{2} ~|~ A = 1)$, and both these terms can be calculated off-line. As a result, we use a detection strategy that measures the MI between $y^{n'}_{1}$ and $y^{n'}_{2}$ by using scikit-learn \cite{Scikit} library's MI calculation method using $k$ nearest neighbors \cite{MI1}. The proposed detection strategy feeds an appropriate value of $\hat{A}$ to the combining block depending on whether the MI estimate is above or below the threshold, which in turn is empirically chosen such that $p_{md}$ is bounded by 0.1\%. 

\textit{2) Neural Network (NN) Classifier:} In this method, we pose attack detection as a supervised classification problem. The proposed NN uses two hidden layers with ReLU activation function followed by a sigmoid output at the end. The inputs to the training phase constitutes channel outputs, namely, $\{y^{n'}_{1}, y^{n'}_{2}\}$ (with 50\% of the frames under attack) along with the respective ground truths on $A$. Based on the inputs, the NN estimates the probability of attack by minimizing an appropriate binary cross-entropy function. We train for eight epochs to ensure convergence over the training set with a batch size of $512$ using the Adam optimizer \cite{adam}. To achieve the constraint of $p_{md} = 0.1\%$, we empirically find an appropriate threshold which gives 0.1\% miss-detection rate on the training data set, and then measure $p_{md}$ and $p_{fa}$ on the validation data set. 

\begin{center}
\begin{figure}
\includegraphics[scale=0.47]{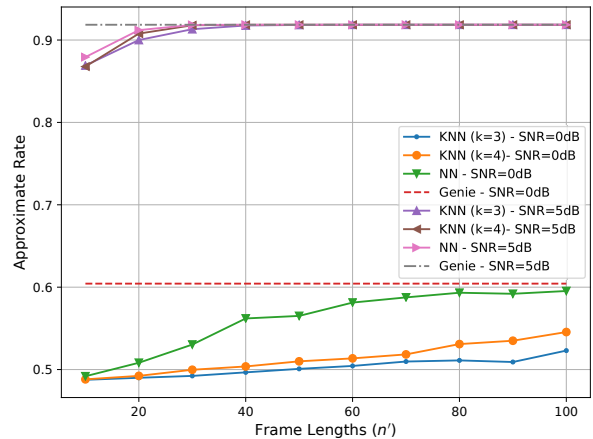}
\vspace{-0.4cm}
\caption{\label{fig:ar}$\mathcal{M}^{approx}_{p_{fa}}$ of attack detectors based on KNN and NN classifier for various $n' \in \{10, 20, \ldots, 100\}$ and SNR $= \{0, 5, 10, 15\}$ in dB. We omit the results for SNR $= 10, 15$ since both detectors achieve the Genie bound.}
\end{figure}
\end{center}

For each combination of $n' \in \{10, 20, \ldots, 100\}$ and $\mbox{SNR} \in \{0, 5, 10, 15\}$, we repeat the experiments to compute $p_{fa}$ of the above detectors by driving their $p_{md} = 0.1\%$. Subsequently, we substitute the corresponding $p_{fa}$ in \eqref{eq:rate_non_genie_pmd_0} to obtain $\mathcal{M}^{approx}_{p_{fa}}$, as presented in Fig. \ref{fig:ar}. The plots show that the NN classifier outperforms the KNN detection significantly at $\mbox{SNR} = 0 \mbox{ dB}$, whereas the benefits of the NN classifier are not significant at $\mbox{SNR} = 5 \mbox{ dB}$. Furthermore, we highlight that $\mathcal{M}^{approx}_{p_{fa}}$ offered by the NN detector is close to that of the Genie detector for frame-lengths as short as $100$ and $40$ symbols at $0$ dB and $5$ dB, respectively. For more details on our experiments, we refer the reader to \cite{github}, where the source codes of the detectors are also available.

\subsection{Discussion on Relevance of Theorem \ref{thm1}}
\label{sec:discussion}
%The result of Theorem \ref{thm1} proves that $\mathcal{M}^{approx}_{p_{fa}}$ is a good approximation on $\mathcal{M}^{non-Genie}_{p_{md}, p_{fa}}$ only under special conditions on $\mathcal{P}$, $\gamma_{1}$, $\gamma_{2}$ and $\sigma^{2}$. However, when these conditions are not satisfied, we cannot evaluate the difference between $\mathcal{M}^{approx}_{p_{fa}}$ and $\mathcal{M}^{non-Genie}_{p_{md}, p_{fa}}$ due to non-availability of the closed-form expression on the latter term. 
For each $n'$ and $\mbox{SNR}$, we can evaluate the tightness of the MI values in Fig. \ref{fig:ar} by first computing $\mathcal{P}$, and then determining an appropriate $\mu'$ such that $\mathcal{P} \subseteq \mathcal{R}_{\mu'}$. With that, \eqref{eq:rate_non_genie_pmd_0} qualifies as an approximation with accuracy $\mu'$. Although obtaining the performance profile $\mathcal{P}$ through exhaustive experiments is computationally challenging for large $n$, sampling techniques can be used to estimate $\mu'$. For instance, at $n' = 50$ and $\mbox{SNR} = 0 \mbox{ dB}$, we have used the NN classifier to empirically compute the pairs $\{(p_{md | \bar{x}}, 1 - p_{fa | \bar{x}})\}$ for $10000$ randomly chosen codewords, and have verified that more than $99\%$ of them lie inside $\mathcal{R}_{\mu'}$ with $\mu' = 3 \times 10^{-3}$. 

As the second caveat, we recollect that Theorem \ref{thm1} is applicable if $\gamma_{1}, \gamma_{2}$ and $\sigma^{2}$ satisfy the conditions in Lemma \ref{lemma2}. However, for arbitrary values of $\gamma_{1}, \gamma_{2}$ and $\sigma^{2}$, we do not have a proof on the applicability of the upper bound in \eqref{pdf_bound1} for all $y^{n} \in \mathbb{R}^{n}$, nor we can verify \eqref{pdf_bound1} for a given $y^{n} \in \mathbb{R}^{n}$ due to intractable distributions. By acknowledging these limitations we caution the reader not to interpret the plots in Fig. \ref{fig:ar} as exact MI values. Nevertheless, we have presented  $\mathcal{M}^{approx}_{p_{fa}}$ as they serve as benchmarks for comparison with tighter approximations on $\mathcal{M}^{non-Genie}_{p_{md}, p_{fa}}$ in future. 

\end{document}